\documentclass[a4paper,UKenglish,cleveref, autoref, thm-restate]{lipics-v2021}

\newcommand{\pace}{\textsf{PACE 2024}}
\newcommand{\oscm}{\texttt{oscm}}
\newcommand{\bac}{branch\&cut}
\newcommand{\coin}{\textsf{coin-or}}
\newcommand{\cbc}{\textsf{Cbc}}

\hideLIPIcs
\nolinenumbers
\relatedversion{To appear in LIPIcs, Volume 321, IPEC 2024}

\title{PACE Solver Description: Exact Solution of the One-sided Crossing Minimization Problem by the MPPEG Team}
\titlerunning{PACE Solver Description by the MPPEG Team}

\author{Michael J{\"u}nger}{University of Cologne, Germany}{juenger-sfb@informatik.uni-koeln.de}{https://orcid.org/0000-0002-6480-2614}{}
\author{Paul J.\ J{\"u}nger}{University of Bonn, Germany}{s94pjuen@uni-bonn.de}{https://orcid.org/0009-0008-4165-4453}{}
\author{Petra Mutzel}{University of Bonn, Germany}{petra.mutzel@cs.uni-bonn.de}{https://orcid.org/0000-0001-7621-971X}{}
\author{Gerhard Reinelt}{Heidelberg University, Germany}{ip121@uni-heidelberg.de}{https://orcid.org/0000-0002-7193-501X}{}

\authorrunning{M.~J{\"u}nger, P.~J.~J{\"u}nger, P.~Mutzel, and G.~Reinelt}

\Copyright{Michael J{\"u}nger, Paul J.~J{\"u}nger, Petra Mutzel, and Gerhard Reinelt} 



\category{PACE Solver Description}

\acknowledgements{The authors gratefully acknowledge the granted access to the Marvin cluster hosted by the University of Bonn.}

\EventEditors{\'{E}douard Bonnet and Pawe\l{} Rz\k{a}\.{z}ewski}
\EventNoEds{2}
\EventLongTitle{19th International Symposium on Parameterized and Exact Computation (IPEC 2024)}
\EventShortTitle{IPEC 2024}
\EventAcronym{IPEC}
\EventYear{2024}
\EventDate{September 4--6, 2024}
\EventLocation{Royal Holloway, University of London, Egham, United Kingdom}
\EventLogo{}
\SeriesVolume{321}
\ArticleNo{27}

\begin{document}

\maketitle

\begin{abstract}
This is a short description of our solver \oscm\ submitted by our team MPPEG to the 
\pace\ challenge both for the exact track and the parameterized track, 
available at \url{https://github.com/pauljngr/PACE2024}~\cite{GitHub_OSCM_2024} 
and \url{https://doi.org/10.5281/zenodo.11546972}~\cite{OCDM-PACE2024-Zenodo}.
\end{abstract}

\section{Method}
\label{sec:method}

We apply the approach to the one-sided crossing minimization problem presented in~\cite{JGAA-1}. 
This article is surveyed by Patrick Healy and Nikola S.~Nikolov in Chapter~13.5 
of the Handbook of Graph Drawing and Visualization~\cite{HealyNikolov} that is 
recommended on the \pace\ web page~\cite{PACE_2024}. The method consists of a transformation of a one-sided 
crossing minimization instance to an instance of the linear ordering problem that is solved by 
\bac\ as introduced in~\cite{GJR1984} and~\cite{GJR1985b}. 
We also use  problem decomposition and reduction 
techniques as well as a heuristic for finding a good initial solution.
With the required brevity, we give a rough sketch of the major details.

The instances of the \pace\ challenge problem consist of a bipartite graph $G=(T\dot\cup B,E)$ 
and a fixed linear ordering $\pi_T=\langle t_1,t_2,\ldots,t_m\rangle$  of $T$ (``the top nodes''). In the 
exact track and the parameterized track, the task is to find a 
linear ordering $\pi_B$ of $B=\{b_1,b_2,\ldots,b_n\}$  (``the bottom nodes'') such that the 
number of edge crossings in a straight-line drawing of $G$ 
with $T$ and $B$ on two parallel lines, following their linear orderings, is provably minimum. 
The NP-hardness of this task has been shown in~\cite{EW1994}. 

For a linear ordering $\pi_B$ of $B$ let
\[x_{ij}=\left\{
\begin{array}{ll}
1&\mbox{if $b_i$ appears before $b_j$ in $\pi_B$,}\\
0&\mbox{otherwise.}
\end{array}
\right.\]
For $i,j\in\{1,2,\ldots,n\}$ let $c_{ii}=0$, and for $i\ne j$ let $c_{ij}$ denote the number of crossings 
between the edges incident to $b_i$ with the edges incident to $b_j$ if $b_i$ appears  
before $b_j$ in $\pi_B$.
Then the number of crossings induced by $\pi_B$ is
\[\mbox{cr}(\pi_B)=\sum_{i=1}^n\sum_{j=1}^nc_{ij}x_{ij}.\]
Since for any pair $b_i\ne b_j$ in $B$ we have $x_{ji}=1-x_{ij}$, we can reduce the number of 
variables to $\binom{n}{2}$ and obtain
\[\mbox{cr}(\pi_B)=\sum_{i=1}^{n-1}\sum_{j=i+1}^{n}c_{ij}x_{ij}+c_{ji}(1-x_{ij})=
\sum_{i=1}^{n-1}\sum_{j=i+1}^{n}(c_{ij}-c_{ji})x_{ij}+\sum_{i=1}^{n-1}\sum_{j=i+1}^{n}c_{ji}.\]
For $a_{ij}=c_{ij}-c_{ji}$ we solve the \emph{linear ordering problem} as
the following binary linear program,
based on the complete digraph $D$ with node set $B$.
\[\begin{array}{rll}
\hbox{(LO)\qquad minimize}&\displaystyle\sum_{i=1}^{n-1}\sum_{j=i+1}^{n}a_{ij}x_{ij}\\
\hbox{subject to} &\displaystyle\sum_{\substack{(b_i,b_j)\in C: \\i<j}}x_{ij}+
\displaystyle\sum_{\substack{(b_i,b_j)\in C: \\i>j}}(1-x_{ji})\le |C|-1&\hbox{for all dicycles $C$ in $D$}\\
&0\le x_{ij}\le 1& \hbox{for } 1\le i<j\le n\\
&x_{ij} \hbox{ integral} & \hbox{for } 1\le i<j\le n.
\end{array}\]
If $z$ is the optimum value of (LO), $z+\sum_{i=1}^{n-1}\sum_{j=i+1}^{n}c_{ji}$ is the minimum number 
of crossings. 
Notice that the classical linear ordering formulation~\cite{GJR1984,GJR1985b} uses 
constraints for cycles of length three only. However, in our approach we also need longer cycles, 
since we remove some of the arcs as we shall describe in Section~\ref{sec:algorithmimplementation}.
The constraints of (LO) guarantee that the solutions correspond precisely to all permutations 
$\pi_B$ of $B$.
Furthermore, it can be shown that for complete digraphs the ``3-cycle constraints'' are necessary in any minimal description 
of the feasible solutions by linear inequalities, if the integrality conditions are dropped. The NP-hardness 
of the problem makes it unlikely that such a complete linear description can be found. 
Further classes of inequalities with a number of members exponential in $n$ that must 
be present in a complete linear description of the feasible set, are known, and some of them can be 
exploited algorithmically. Indeed, small M\"obius-ladder constraints, the one shown in Figure~3 of~\cite{GJR1984},
as well as the same in which all arcs are reversed, have been found useful in this
crossing minimization context. 

\section{Algorithm and Implementation}
\label{sec:algorithmimplementation}

When the integrality conditions in (LO) are dropped,  we obtain a linear programming relaxation of (LO) which has been proven 
very useful in practical applications. The structure of our \bac\ algorithm \oscm\ 
(``\textbf{o}ne-\textbf{s}ided \textbf{c}rossing \textbf{m}inimization'') is similar to the one 
proposed in~\cite{GJR1984}.
The algorithm starts with the trivial constraints $0\le x_{ij}\le1$
that are handled implicitly by  the linear programming solver, iteratively adds violated cycle and M\"obius-ladder constraints, 
and deletes 
nonbinding constraints after a linear program has been solved, until the relaxation is solved. This requires a 
\emph{separation} algorithm that, given the solution of some relaxation, is able to determine a violated 
inequality called \emph{cutting plane}.
If the optimum solution of the relaxation is integral, the algorithm stops, otherwise it is applied recursively to two 
subproblems in one of which a fractional $x_{ij}$ is set to $1$ and in the other set to $0$. Thus, in the end,
an optimum solution is found as the solution of some relaxation, along with a proof of optimality.

\oscm\  makes use of the following observations, some of which stem from the literature in fixed-parameter 
algorithms for one-sided crossing minimization. 
Lemma~\ref{lemma:decompose}
allows us to decompose the given instance. Within the components, we can fix and eliminate
variables from (LO) by
Lemma~\ref{lemma:fix}, and we can exclude variables $x_{ij}$ with $a_{ij}=0$ from (LO) by Lemma~\ref{lemma:arbitrary}.

\begin{lemma}[Decomposition]\label{lemma:decompose}
For each node $v\in B$, we define the open interval $I_v=]l_v,r_v[$, where $l_v$ is the position of the 
leftmost and $r_v$ the position of the rightmost neighbor of $v$ in $\pi_T$. 
The union of the intervals $I_v$ 
induces a partition $B_1,B_2,\ldots,B_k$ of $B$ such that every $I_{B_i}=\bigcup_{v\in B_i} I_v$, $i=1,\ldots,k$, is an 
interval, and for any pair  $B_i,B_j$ the intervals $I_{B_i}$ and $I_{B_j}$ are disjoint. 
In every optimum $\pi_B$ all the nodes of $B_i$ appear before those 
of $B_j$ if $I_{B_i}$ is to the left of $I_{B_j}$.
\end{lemma}

Indeed, $51$ of the $100$ exact-public instances have 
between $2$ and $154$ components.
\begin{lemma}[Variable fixing~\cite{DBLP:journals/algorithmica/DujmovicW04}]\label{lemma:fix}
If for any pair of nodes $b_i,b_j\in B$, we have $c_{ij}=0$ and  $c_{ji}>0$, then every optimal solution 
of (LO) satisfies $x_{ij}=1$, if $i<j$, and $x_{ji}=0$, if $i>j$.
\end{lemma}

\begin{lemma}[Arbitrary ordering]\label{lemma:arbitrary}
Let $\pi_B^{(p)}$ be a partial ordering induced by the variables $x_{ij}$ with $a_{ij}\not=0$, then there exist values
$x_{ij}\in\{0,1\}$ for $a_{ij}=0$ defining a total ordering $\pi_B$ of $B$ with no effect to the objective function value.
This assignment can be found by topologically sorting $B$ with respect to $\pi_B^{(p)}$.
 \end{lemma}
 
This setup has the advantage that (sometimes considerably) smaller linear 
programs need to be solved, but, on the other hand, separation becomes more involved. 
In order to obtain an optimal partial ordering $\pi_B^{(p)}$ of $B$ using the variables left in (LO), 
we need to include cycle constraints for larger cycles as already mentioned in Section~\ref{sec:method}.

For computational efficiency, \oscm\ has a hierarchy of separation procedures. The first for 3-dicycles 
 is based on depth first search. The second for dicycles of length at least $4$  with 
integral weights is also based on depth first search. Violated dicycles are shortened via breadth first 
search, restricted to the cycle nodes,  starting from back arcs of the preceding depth first search. 
The third applies shortest path techniques for separation of cycles containing fractional arcs as described 
for the related acyclic subdigraph problem in section~5 of~\cite{GJR1985a}. 
First, the above separation procedures are applied on the graph containing only the arcs present in (LO).
If all of the above do not find any violated inequalities, \oscm\ extends the search to the fixed arcs. 
After separation, the linear program is resolved using the dual simplex method providing the same or a better 
lower bound on the minimum number of crossings. If the progress 
compared to the previous bound is small for a sequence of such lower bounds,  
\oscm\ applies a 
heuristic for finding violated M\"obius ladder inequalities, and if this does not lead to a significant 
improvement, the \bac\ phase is started.

Whenever a linear program has been solved, it is checked by topological sorting if the solution is 
the characteristic vector of a linear ordering. If not, a relaxed topological sorting procedure is applied 
in the pursuit of finding a better incumbent solution that provides an upper bound for the minimum 
number of crossings. \oscm\ stops when the (integral) upper bound and the (possibly fractional) 
lower bound differ by less than~$1$, proving optimality.

For small instances, \oscm\ applies a variant of the heuristic 
``Kernighan-Lin~2'' of~\cite{martireinelt2022} for finding a decent initial solution before the 
optimization starts.

\section{Performance}
\label{sec:performance}

Our program \oscm, published in~\cite{GitHub_OSCM_2024} and~\cite{OCDM-PACE2024-Zenodo}, consists 
of roughly $3500$ lines of C/C++ code. It makes use of the \coin~\cite{coin} \cbc\ library, version 2.10.7~\cite{cbc}.

We submitted \oscm\ both to the exact track and the parameterized track of \pace. In the official ranking, \oscm\ 
received the first place in the exact track with 199 of the 200 instances instances solved in about 5682 seconds, 
and the third place in the paramaterized track with all 200 instances solved in about 25 seconds.

\bibliography{pacesolverdescriptionMPPEG}

\end{document}